# RENTES EN COURS DE SERVICE : UN NOUVEAU CRITÈRE D'ALLOCATION D'ACTIFS

Frédéric PLANCHET[*]     Pierre-E. THÉROND[α]

ISFA – Université Lyon 1 [β]
Winter & Associés [γ]

RÉSUMÉ

Nous comparons deux méthodes d'allocation initiale d'actifs d'un régime de rentes en cours de service dans le cas simplifié du choix de portefeuille entre un actif localement sans risque dont le rendement est modélisé par un processus CIR et un actif risqué dont le cours est modélisé par un mouvement brownien géométrique. Les deux méthodes d'allocation d'actifs proposées se fondent sur les critères de probabilité de ruine et de maximisation des fonds propres économiques. Nous abordons successivement les cas où les flux de passif sont connus puis aléatoires, ce qui nous permet d'étudier la décomposition du risque entre risque démographique (mutualisable) et risque financier (non mutualisable) puis l'impact de la prise en compte de la nature aléatoire des flux de prestations sur le choix de portefeuille. Enfin nous introduisons un processus de revalorisation des rentes lié à l'évolution de l'inflation et observons son impact sur les allocations obtenues par les deux critères retenus.

MOTS-CLEFS :   Régime à prestations définies, allocation d'actifs, probabilité de ruine, fonds propres économiques, mutualisation, revalorisation de rentes.

ABSTRACT

The aim of this paper is to compare two asset allocation methods for a pension scheme during the decumulation phase in the simplified portfolio selection between a risky asset following a geometric Brownian motion and a riskless asset. The two asset allocation criteria are the ruin probability of the insurance company and the optimization of the economic capital. We first solve the asset allocation problem with deterministic pension payments then with stochastic mortality risk. We analyze the part of mortality risk in the global risk of the company. Then we show the impact of the indexation of the pensions to the inflation on the asset allocation.

*Journal of Economic Literature Classification: G11, G23.*

# 1. INTRODUCTION

La détermination de l'allocation d'actif est un thème central dans les problématiques d'assurance vie, particulièrement développé dans le contexte des contrats d'épargne et des régimes de retraite supplémentaire. Dans le cas de la retraite, ou de prestations de rentes viagères, la durée des contrats permet aux services de gestion actif/passif d'élaborer des stratégies d'allocation d'actifs à long terme *a priori* indépendantes de spéculations à court terme (qui relèvent de l'allocation tactique).

La gestion d'un régime de retraite fait intervenir de ce point de vue deux phases distinctes : une première phase de constitution des droits, avec la distinction classique entre les régimes à cotisations définies et les régimes à prestations définies, et une seconde phase de restitution, une fois la rente liquidée. Cette distinction est particulièrement marquée dans le cas des régimes d'entreprise assurés, la nature de l'engagement de l'assureur étant complètement différente dans chacune de ces deux phases ; en pratique l'assureur n'assume en effet un engagement viager que pour la phase de restitution, après le prélèvement du capital constitutif de la rente du fonds des actifs pour le transférer dans le fonds des retraités[1] (l'insuffisance du fonds en phase de constitution conduit simplement à un appel de cotisation complémentaire). Incidemment, il faut donc noter que les fonds des actifs et des retraités sont dans ce cas distincts et les allocations sont différentes dans ces deux phases (cette question est abordée par exemple par BATTOCHIO et al. [2004]).

Nous nous intéressons dans ce travail à la question de l'allocation d'actifs en phase de restitution uniquement. En particulier, nous n'abordons pas la question de la fixation du niveau du taux de cotisation d'un régime à prestations définies, qui est une problématique connexe à celle de l'allocation d'actifs pour le fonds des actifs ; une abondante littérature est consacrée à ce sujet : on pourra par exemple consulter les travaux de BOULIER et al. [2001b], JOSA-FOMBELLIDA et RINCÓN-ZAPATERO [2004]). L'analyse de l'engagement de tels régimes d'entreprise est effectuée par MAGNIN et PLANCHET [2000].

L'analyse de la dynamique du régime en phase de restitution conduit dans un premier temps à considérer le groupe fermé des pensionnés en cours de service. Lorsque le régime n'est plus en phase de montée en charge, l'extension au groupe ouvert peut être ensuite prise en compte au travers du pilotage du régime, qui conduit à actualiser régulièrement les études actuarielles sur la base des données révisées. Ce mécanisme inclut de fait une auto-correction et un ajustement lissé au fil du temps des paramètres techniques du régime. Cette approche est de plus en phase avec la gestion de ce type de régime par les assureurs[2]. Elle est détaillée dans PLANCHET et al. [2005].

---

[1] On peut d'ailleurs remarquer que le distinction cotisations définies / prestations définies est dans ce contexte décisive pour l'entreprise, mais pas pour l'assureur d'un régime supplémentaire de retraite.
[2] Ce point sera repris dans la suite de ce travail.

Les premiers modèles d'allocation d'actifs à intégrer le risque lié aux placements ont été inspirés de techniques financières, en particulier de critères de type MARKOWITZ [1952]. Mais ces modèles n'intègrent en général pas les contraintes propres à un régime de rente. Certains auteurs tels que LEIBOWITZ et KOGELMAN [1991] et CAMPBELL et al. [2001] ont proposé des adaptations de ces modèles pour intégrer ces contraintes spécifiques et, en particulier, la prise en compte de la probabilité de réaliser un rendement minimum (shortfall-risk). Toutefois, BOYLE [2004] montre les limites de l'utilisation de telles approches lorsque l'on intègre l'estimation des paramètres des actifs, qui peut conduire à une allocation sous-optimale.

Depuis 2001, des auteurs comme BATTOCHIO et al. [2004], CHARUPAT et MILEVSKY [2002] ou encore BOULIER et al. [2001a] ont développé des modèles intégrant les contraintes assurantielles (notamment le risque de mortalité), mais leur mise en œuvre pratique est délicate du fait de la problématique du choix d'une fonction d'utilité et de l'estimation d'un nombre important de paramètres. Une synthèse des travaux récents sur le sujet est proposée dans GRASSELLI [2004].

Par ailleurs, dans le cadre des réflexions sur la solvabilité des organismes assureurs issues de « solvabilité 2 »[3], de nouveaux modèles intégrant les paramètres de solvabilité par le biais d'une contrainte sur la probabilité de ruine sont apparus. On détermine ainsi une allocation d'actifs qui contrôle la probabilité de ruine de l'assureur ou en d'autres termes la capacité de faire face à ses engagements, au travers de critères de type Value-at-Risk ou Tail-Value-at-Risk. DJEHICHE et HORFELT [2004] proposent une synthèse des différents modèles candidats à devenir des « modèles standards ».

Dans ce contexte, l'objectif de ce travail est de proposer une démarche de détermination de l'allocation stratégique spécifique des régimes de rentes en phase de restitution intégrant les particularités de l'assurance, sans nécessité de fixer *a priori* un paramètre exogène tel que le niveau de la probabilité de ruine. L'objectif poursuivi est de proposer un critère d'allocation pertinent et opérationnel. Afin de focaliser l'exposé sur le choix du critère proposé et ses propriétés, un certain nombre d'hypothèses simplificatrices sont effectuées ; elles seront relâchées dans de futurs développements.

Par ailleurs, nous choisissons délibérément une modélisation très agrégée du régime considéré, suivant en cela la démarche proposée par JANSSEN [1992] et DEELSTRA et JANSSEN [1998] : nous disposons au passif d'un échéancier de flux (aléatoire) et à l'actif de deux supports, l'un (localement) sans risque et l'autre risqué. Nous n'abordons en particulier pas la modélisation du mécanisme de fixation du niveau des rentes, défini par le règlement intérieur du régime. Notre problématique se situe en aval.

---

[3] Voir notamment les travaux de l'AAI [2004] et ceux de la COMMISSION EUROPÉENNE [2004].

## 2. CARACTÉRISTIQUES DU PORTEFEUILLE DE RENTES

Dans la suite, nous reprenons pour les applications numériques l'exemple utilisé dans PLANCHET et THÉROND [2004]. Il s'agit d'un portefeuille constitué de 374 rentiers âgés en moyenne de 63,8 ans au 31/12/2003. La rente annuelle moyenne s'élève à 5 491 €. Le graphique 1 présente les flux de prestations espérés en fonction du temps obtenus à partir de la table de mortalité TV 2000.

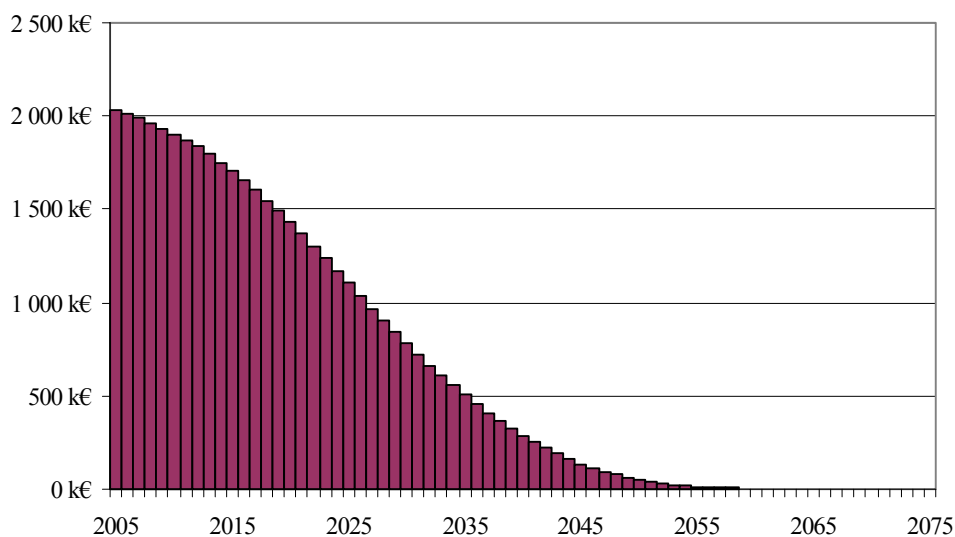

Fig. 1 :   Flux de prestations espérés

Avec un taux d'escompte des provisions de 2,5 %, la provision mathématique initiale, s'élève à 32,8 M€. La duration[4] du passif est alors de 12,3 ans.

On notera que la courbe ci-dessus est essentiellement en pratique non paramétrique : sa détermination passe par l'application aux flux de prestations, connus, d'une table de mortalité exprimée sous une forme non paramétrique. La table de mortalité étant dans ce contexte en général prospective, les méthodes d'ajustement à des lois paramétriques classiques, telles que le modèle de Makeham par exemple (*cf.* PETAUTON [2004]), ne s'appliquent pas. Toutefois, un ajustement polynomial peut être effectué pour disposer d'une formulation paramétrique *ex post* :

---

[4] On rappelle la définition de la duration de Macaulay, qui peut-être interprétée comme la durée de vie moyenne d'un échéancier de flux : $\mathbf{D}(f;i) = \frac{1}{V} \sum_{k \geq 1} k * f_k * (1+i)^{-k}$ où $V = \sum_{k \geq 1} f_k * (1+i)^{-k}$.

- 4 -

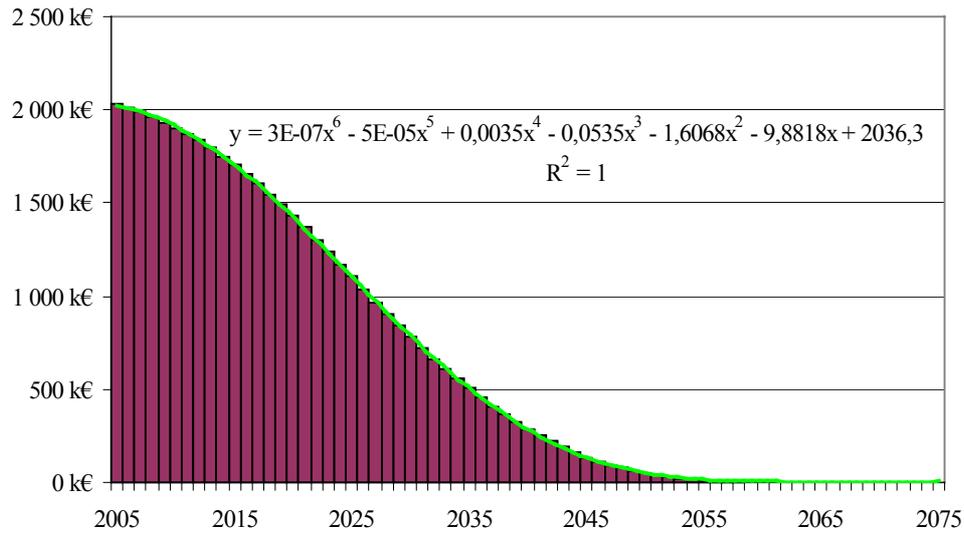

Fig. 2 :   Flux de prestations espérés ajustés

Cette paramétrisation du modèle ne simplifie toutefois pas sensiblement les calculs par la suite.

## 3. LE MODÈLE

Dans la suite de cet article, $E_t$ désignera le montant des fonds propres à la date $t$. Celui-ci sera obtenu comme étant la différence entre le montant de l'actif $A_t$ et le montant des provisions mathématiques $L_t$. Si l'on note $\tilde{F}_t$ le flux de prestation aléatoire qui aura lieu à la date $t$, la provision mathématique $L_t$ s'obtiendra comme étant la valeur actuelle probable de ces flux au taux d'escompte discret $i$.

Nous noterons **J** l'ensemble des individus et $x(j)$ désignera l'âge en 0 de l'individu $j$ tandis que $f_j$ sera le montant de sa rente annuelle.

Enfin, nous nous placerons systématiquement dans l'univers historique par le biais de la probabilité historique **P** et de la filtration **Φ** associée à l'information disponible.

### 3.1.   DYNAMIQUE DU BILAN DE L'ASSUREUR

Le portefeuille est exclusivement constitué de rentes en cours de service supposées non réversibles. L'assureur ne reçoit donc plus de primes et doit donc gérer au mieux ses actifs afin de pouvoir honorer son engagement vis-à-vis des bénéficiaires.

En 0, il estime la suite de flux probables de prestations que nous noterons $(F_t)_{t \geq 1}$ :

$$F_t = \mathbf{E}\left[\widetilde{F}_t \mid \Phi_0\right], \tag{1}$$

où

$$\widetilde{F}_t = \sum_{j \in \mathbf{J}} f_j * \mathbf{1}_{]t;\infty[}(T_{x(j)}), \tag{2}$$

où $T_{x(j)}$ désigne la date de décès (aléatoire) de la tête d'âge $x(j)$.

Avec les notations classiques de l'assurance vie, où $l_x$ désigne le nombre de personne en vie à l'âge $x$ sur une base 100 000 à l'âge 0 ($l_0 = 100000$), on a :

$$F_t = \sum_{j \in \mathbf{J}} f_j * \frac{l_{x(j)+t}}{l_{x(j)}}. \tag{3}$$

A partir de cette estimation, la provision mathématique initiale $L_0$ s'obtient par l'actualisation au taux d'intérêt technique $i$ de ces flux :

$$L_0 = \sum_{t=1}^{\infty} F_t (1+i)^{-t}. \tag{4}$$

Cette approche, issue de la réglementation française, est entièrement compatible avec les normes IFRS puisque la seule différence dans le calcul de la provision mathématique réside dans le choix du taux d'intérêt technique (*cf.* 4.2.1 *infra*).

Le bilan de l'assureur en 0 peut alors s'écrire :

**Bilan en 0**

| | |
|---|---|
| $A_0$ | $E_0$ |
| | $L_0$ |

Supposons que les prestations sont servies en début d'année, le bilan évolue alors selon le processus suivant :

$$\begin{cases} L_t = \sum_{k=t+1}^{\infty} \frac{\mathbf{E}\left[\widetilde{F}_k \mid \Phi_t\right]}{(1+i)^{k-t}} \\ A_t = (1+\widetilde{R}_t)A_{t-1} - \widetilde{F}_t \\ E_t = A_t - L_t \end{cases} \tag{5}$$

où $\widetilde{R}_t$ désigne la rentabilité (aléatoire) du portefeuille financier entre les dates $t-1$ et $t$.

## 3.2. LE MARCHÉ FINANCIER

L'assureur doit constituer son portefeuille financier à partir d'un actif localement sans risque et d'un actif risqué. Nous supposerons que l'actif localement sans risque $Y$ a un rendement instantané $r_t$ qui évolue selon le modèle de Cox, Ingersoll et Ross (CIR) :

$$dr_t = a(r_\infty - r_t)dt + \sigma_r \sqrt{r_t} dB_t^r, \tag{6}$$

où $B^r$ est un **P**-mouvement brownien. Nous supposerons par ailleurs, que l'actif risqué $X$ suit un mouvement brownien géométrique

$$\frac{dX_t}{X_t} = \mu \, dt + \sigma_X \, dB_t^X, \tag{7}$$

où $B^X$ est un **P**-mouvement brownien dont la dépendance avec $B^r$ est linéaire et peut être résumée par la corrélation

$$dB_t^r dB_t^X = \rho dt. \tag{8}$$

De manière à ce que le processus (6) soit bien défini pour $t \in \mathbf{R}_+$, nous choisirons des paramètres qui respectent l'inégalité

$$a \geq \frac{\sigma_r^2}{2}. \tag{9}$$

Celle-ci assure que $\mathbf{Pr}\left[\exists t < +\infty \mid r_t \leq 0\right] = 0$ (*cf.* LAMBERTON et LAPEYRE [1997] pour la démonstration de ce résultat).

Pour simplifier les écritures, on posera, sans perte de généralité : $X_0 = Y_0 = 1$. Nous supposerons également que :

$$\mu \geq r_\infty \geq \mathbf{ln}(1+i) \geq 0. \tag{10}$$

Cette hypothèse est naturelle puisque le rendement espéré d'un actif risqué doit être plus élevée que celui d'un placement peu risqué et que le taux maximal d'escompte des provisions, qui est fixé réglementairement, est, par prudence, moins élevé que le rendement escompté des actifs.

L'évolution conjointe des processus $X$ et $r$ sera simulée à partir d'une discrétisation exacte du mouvement brownien géométrique et d'une discrétisation selon le schéma de Milstein pour

le processus CIR (*cf.* PLANCHET et al. [2005] pour une description des différents procédés de discrétisation).

### 3.3. LA GESTION DES ACTIFS

Lorsque l'assureur est amené à désinvestir pour payer les prestations, nous supposerons qu'il vend les deux actifs de manière proportionnelle à leur part respective dans la valeur de marché du portefeuille. De manière pratique, cela revient à considérer que l'actif est investi en 0 dans un fonds qui ne fera plus l'objet d'arbitrage et dont l'assureur vendra des parts pour régler les rentes.

Cette hypothèse est moins restrictive qu'il n'y paraît en première lecture : en effet, en pratique l'étude conduisant à déterminer l'allocation optimale sera effectuée chaque année et une nouvelle allocation sera alors déterminée ; ce mécanisme permet notamment d'intégrer « au fil de l'eau » les corrections des erreurs de prédiction sur le niveau de la mortalité. La prise en compte année après année des décès constatés est un point très important, la déformation de la charge future n'étant pas seulement due aux aspects démographiques (*i. e.* le nombre de décès par âge), mais également au niveau des prestations servies aux bénéficiaires décédés (deux bénéficiaires de même âge pouvant avoir des prestations très différentes).

Cette approche est justifiée par la grande inertie du régime : on cherche alors chaque année la trajectoire optimale compte tenu de la situation de la population d'une part et des marchés financiers d'autre part. En ce sens, la prise en compte de réallocations annuelles fixées *a priori* complique le modèle inutilement et nous préférons les intégrer *a posteriori* par le pilotage du régime. Cette démarche est bien connue des assureurs vie notamment dans le cadre des contrats collectifs pour les régimes d'indemnités de fin de carrière ou de retraite supplémentaire, pour lesquels le taux de cotisation d'équilibre est déterminé chaque année sur la base de l'information disponible, comme s'il allait être constant jusqu'à l'extinction du groupe.

Formellement, on a alors :

$$\begin{aligned} A_1 &= \left(\theta X_1 + (1-\theta)Y_1\right)A_0 - F_1 = \left(\theta X_1 + (1-\theta)Y_1\right)\left[A_0 - \frac{F_1}{\theta X_1 + (1-\theta)Y_1}\right] \\ A_2 &= \left(\theta X_2 + (1-\theta)Y_2\right)\left[A_0 - \frac{F_1}{\theta X_1 + (1-\theta)Y_1}\right] - F_2 \\ A_2 &= \left(\theta X_2 + (1-\theta)Y_2\right)\left[A_0 - \sum_{s=1}^{2} \frac{F_s}{\theta X_s + (1-\theta)Y_s}\right] \end{aligned} \quad (11)$$

où θ désigne la proportion initialement investie en actif risqué. Par récurrence, il vient :

$$A_t = \left(\theta X_t + (1-\theta)Y_t\right)\left[A_0 - \sum_{s=1}^{t}\frac{F_s}{\theta X_s + (1-\theta)Y_s}\right] \quad (12)$$

Notons que cette expression n'est pas équivalente à :

$$A_{t+1} = \left(\theta\frac{X_{t+1}}{X_t} + (1-\theta)\frac{Y_{t+1}}{Y_t}\right)A_t - F_{t+1} \quad (13)$$

Cette approche alternative (13) correspond à la situation dans laquelle l'assureur recompose, chaque début de période, son actif en investissant $\theta$ en actif risqué $X$ et $1-\theta$ en actif localement sans risque $Y$.

Le problème auquel se trouve confronté l'assureur est de composer en 0 son portefeuille d'actifs de manière optimale. Nous verrons dans la suite ce qu'il faut entendre par « de manière optimale ».

Dans la suite du développement, nous supposerons que les suites de flux de prestations et de rendements des actifs sont indépendantes. Cette hypothèse, assez naturelle dans une situation standard, peut s'avérer fausse dans des situations extrêmes telles qu'une guerre civile qui risque de voir conjointement la mortalité exploser et le cours des valeurs mobilières chuter. Nous ne considérerons pas ici ces situations exceptionnelles.

## 4. FLUX DE PRESTATIONS DÉTERMINISTES

Supposons que les flux de prestations sont connus en 0. Cette situation correspond au cas d'un portefeuille important, condition qui n'est pas réalisée dans notre exemple (le cas général est traité au paragraphe 5).

Cela signifie que les effectifs à chaque âge sont suffisamment importants pour que l'approximation suivante soit validée :

$$\widetilde{F}_t \approx \mathbf{E}\left[\widetilde{F}_t\right] = F_t. \quad (14)$$

D'après (5), le bilan évolue alors de la manière suivante :

$$\begin{cases} L_{t+1} = (1+i)L_t - F_{t+1} \\ A_{t+1} = (1+\widetilde{R}_{t+1})A_t - F_{t+1} \\ E_{t+1} = A_{t+1} - L_{t+1} \end{cases} \quad (15)$$

Notons que *E* est alors une sous-martingale. En effet d'après (15), il vient :

$$E_{t+1} = (1 + \widetilde{R}_{t+1})E_t + (\widetilde{R}_{t+1} - i)L_t, \qquad (16)$$

or la condition (10) assure que

$$\forall t \geq 0, \mathbf{E}\left[\widetilde{R}_{t+1} \mid \Phi_t\right] \geq i \geq i \frac{L_t}{L_t + E_t}, \qquad (17)$$

donc

$$\mathbf{E}\left[E_{t+1} \mid \Phi_t\right] \geq (1+i)E_t \geq E_t. \qquad (18)$$

Cela signifie que l'actionnaire verra en moyenne les fonds propres augmenter au cours du temps. En effet, du fait du choix prudent du taux d'actualisation, les fonds propres de l'assureur augmenteront en moyenne du fait des produits financiers qu'ils génèrent et, en l'absence de revalorisation des rentes, du fait de la sur-performance espérée des actifs en représentation des engagements techniques.

### 4.1. PROBABILITÉ DE RUINE

Dans cette partie, la contrainte du programme d'optimisation de l'assureur est de contrôler la probabilité de ruine de la société. Il y a ruine de l'assureur lorsque les fonds propres sont réduits à 0, ou en d'autres termes, lorsque l'actif de la société ne suffit plus à couvrir les provisions techniques. Notons τ l'instant de ruine de l'assureur défini comme suit :

$$\tau = \mathbf{Inf}\left\{t \in \mathbf{N} \mid E_t < 0\right\}. \qquad (19)$$

τ est un temps d'arrêt par rapport à la filtration **Φ** puisque c'est le temps d'entrée de la suite $(E_n)$ dans l'ensemble $]-\infty;0[$. La loi de τ est déterminée par la dynamique (15) ; toutefois les expressions non paramétriques de la suite des flux $(F_t)$ d'une part et de la table de mortalité d'autre part ne permettent pas d'obtenir une expression analytique de cette loi, ce même dans le cas où un ajustement polynomial de $(F_t)$ est effectué. La distribution de τ peut néanmoins être étudiée à l'aide de techniques de Monte Carlo.

Le profit espéré provenant du régime de rentes étant croissant avec la rentabilité espérée du portefeuille financier et donc avec la part investie en actif risqué, l'assureur va composer son portefeuille en fonction de la probabilité de ruine qu'il est prêt à accepter. Le programme d'optimisation, que l'assureur doit résoudre, peut donc s'écrire :

$$\mathbf{Sup}\left\{\theta \in [0;1] \mid \mathbf{P}(\tau_\theta < \infty) \leq \pi_{\max}\right\}, \qquad (20)$$

où $\pi_{max}$ désigne la probabilité de ruine maximale que l'assureur peut accepter.

Les techniques de simulations permettent de résoudre simplement ce programme d'optimisation. La part investie en actif risqué en 0 étant indépendante du cours des deux actifs dans le futur, il est possible d'utiliser ces trajectoires pour différentes allocations initiales de manière à pouvoir comparer, sur les mêmes bases, les différentes allocations.

Dans notre cas, si $N$ est le nombre de trajectoires simulées des actifs, en notant $e_k^n(\theta)$ le montant des fonds propres de la société à la date $k$, dans l'état du monde $n$, pour l'allocation initiale définie par $\theta$, un estimateur empirique de la probabilité de ruine est :

$$1 - \frac{1}{N}\sum_{n=1}^{N}\prod_{k\geq 1} I_{[0;\infty[}\left(e_k^n(\theta)\right) \qquad (21)$$

Le lecteur pourra se référer à PLANCHET et JACQUEMIN [2003] pour la génération de réalisations de variables aléatoires[5] et à PLANCHET et al. [2005] pour la discrétisation de processus continus.

Pour notre exemple, nous avons généré 10 000 trajectoires de l'actif. En faisant varier de 0 % à 100 % avec un pas de 0,05 % la part initiale d'actif risqué, nous avons disposé de $2.10^7$ « scénarios » qui nous ont permis de tracer le graphique 3.

Pour les illustrations numériques, nous avons utilisé les paramètres[6] :

$$E_0 = 4\% * L_0 \quad \mu = \ln(1+0,07) \quad \sigma_X = 0,25$$

$$i = 0,025 \quad r_0 = \ln(1+0,03) \quad r_\infty = \ln(1+0,0462)$$

$$\rho = -0,1 \quad a = 0,5 \quad \sigma_r = 0,02$$

Le graphique suivant représente la probabilité de ruine en fonction de la part d'actif risqué du portefeuille financier à l'origine.

---

[5] Nous avons utilisé ici un générateur pseudo-aléatoire pour générer des réalisations de variables aléatoires (v. a.) de loi uniforme qui ont été transformées par la technique de Box-Muller en réalisations de v. a. de loi normale.
[6] Le taux d'intérêt à long terme correspond au taux long terme de la courbe des taux au 31 décembre 2003 publiée par l'Institut des Actuaires (modèle de Vasicek & Fong).

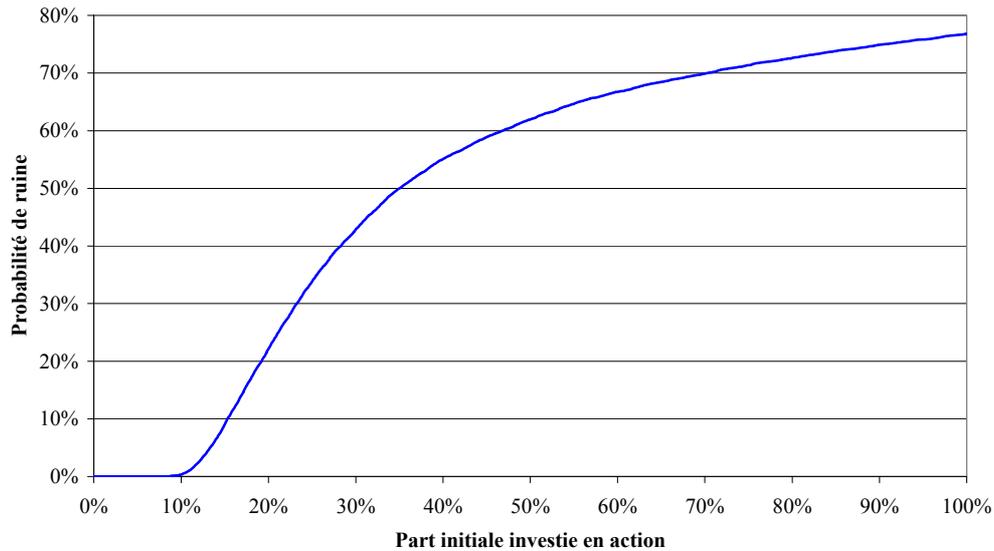

Fig. 3 : Probabilité de ruine en fonction de la part d'actif risqué

Rappelons que la ruine présentée ici s'entend sur la durée totale du régime soit sur plus de 70 ans et que les probabilités de ruine présentées ne peuvent pas être directement comparées aux probabilités de ruine sur un an évoquées dans les projets concernant le futur référentiel prudentiel européen « Solvabilité 2 » (*cf.* COMMISSION EUROPÉENNE [2004]).

Par ailleurs, la ruine définie par l'expression (19) est une « ruine comptable » : on regarde à chaque date de bilan si la valeur des placements est supérieure au montant des provisions calculées avec un taux d'escompte prudent. Par opposition, on pourrait s'intéresser à la probabilité de « ruine économique » qui serait la probabilité que l'assureur ne puisse plus payer les rentes. Le premier instant de ruine économique $\tau'$ serait donc défini comme suit :

$$\tau' = \mathbf{Inf}\left\{t \in \mathbf{N} \mid A_t \leq 0\right\}. \tag{22}$$

En l'absence de précision supplémentaire, c'est de la « ruine comptable » qu'il s'agira lorsque, par la suite, on parlera de ruine.

Dans notre exemple, si l'on souhaite limiter à 1 % la probabilité de ruine du régime, on investira moins de 10,9 % de son actif initial en actif risqué. Cette proportion apparaît relativement faible comparée aux pratiques des sociétés d'assurance sur la vie.

## 4.2. MAXIMISATION DES FONDS PROPRES ECONOMIQUES

### 4.2.1. Contraintes réglementaires

Les dispositions réglementaires françaises actuelles imposent de nombreuses contraintes aux assureurs. Parmi celles-ci les contraintes quantitatives d'escompter les flux futurs dans le calcul de la provision mathématique au taux maximal[7] de **Min** $\{60\% * \text{TME} ; 3,5\%\}$ et de disposer d'un niveau minimal[8] de fonds propres proportionnel à la provision mathématique.

Le graphique 4 reprend l'évolution du TME et des taux réglementaires qui y font référence.

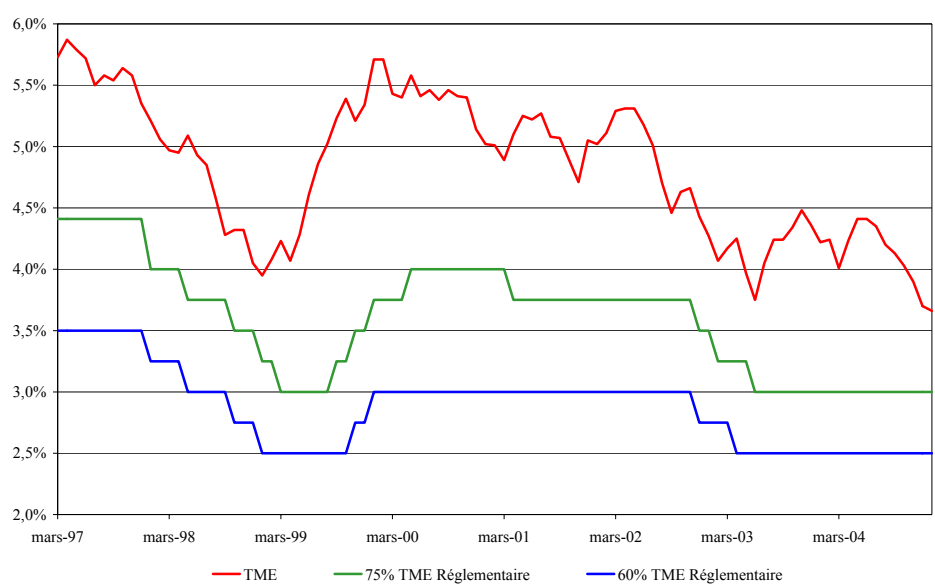

Fig. 4 : Evolution du TME et des taux réglementaires de mars 1997 à janvier 2005

Dans le cadre d'une comptabilisation selon les normes IAS/IFRS, la norme IFRS 4 *Contrats d'assurance* prévoit que les sociétés pourront escompter leurs provisions sur la base d'un taux de marché (le taux sans risque de maturité égale à la duration du passif). On pourra consulter sur ce sujet DOUARD [2000] et VÉRON [2003]. L'application de cette norme conduirait donc à des provisions moins importantes. Toutefois on observera que la norme IFRS 4 ne constitue pas un référentiel prudentiel, mais un référentiel comptable ; les problématiques de solvabilité sont en cours de discussion[9] au niveau européen au travers du dispositif Solvabilité 2 (*cf.* notamment COMMISSION EUROPÉENNE [2002a] et [2002b]). La mise en place du nouveau dispositif prudentiel ne sera pas effective avant 2007, et la forme définitive de celui-ci n'est pas arrêtée ; c'est donc sur la base des comptes établis selon la réglementation française actuelle que la Commission de contrôle des assurances (CCAMIP) continue d'exercer son contrôle prudentiel, et nous retiendrons donc ces règles dans ce travail. Au surplus, bien que selon l'IFRS 4, le niveau des provisions soit moins important que selon la réglementation française, la différence sera intégrée aux fonds propres, ce qui ne modifie pas fondamentalement notre problématique puisque nous ne distinguons pas les actifs en contrepartie des provisions et ceux en contrepartie des capitaux propres.

---

[7] Par prudence, la réglementation française (*cf.* article A.331–10 du Code des assurances) impose à l'assureur d'escompter ses provisions à un taux inférieur à Min (60 %*TME ; 3,5 %), où TME est le Taux Moyen des emprunts de l'Etat français.

[8] Le niveau des fonds propres initiaux est le minimum imposé par la marge de solvabilité en France : 4 % des provisions mathématiques dans notre cas. *Cf.* art. R.334–1 et suivants du Code des assurances.

[9] Les documents sur ce sujet sont disponibles sur le site de la Commission Européenne http://europa.eu.int/comm/internal_market/insurance/solvency_fr.htm

Dans le cadre d'un contrat sans revalorisation des rentes, l'actionnaire tire son profit de deux sources : les revenus financiers engendrés par le capital et le résultat provenant de l'opération d'assurance à proprement parler. En effet, si le taux technique est inférieur au rendement réel des actifs, les provisions mathématiques vont engendrer des profits, le rendement des actifs étant supérieur à l'effet de désactualisation.

### 4.2.2. Fonds propres économiques

Notons $\Lambda_\theta$ la variable aléatoire :

$$\Lambda_\theta = \sum_{t=1}^{\infty} \frac{F_t}{\theta X_t + (1-\theta) Y_t}. \tag{23}$$

Cette variable aléatoire est la valeur limite de la quantité utilisée dans le membre de droite de l'équation (12). $\mathbf{E}[\Lambda_\theta]$ peut s'interpréter comme la provision mathématique « économique » du régime de rentes, *i.e.* la valeur probable des flux futurs actualisés au taux de rendement du portefeuille financier. En effet, si l'on note $\widetilde{R}_t(\theta)$ le rendement du portefeuille d'actifs entre les dates $t-1$ et $t$, il vient :

$$\Lambda_\theta = \sum_{t=1}^{\infty} \frac{F_t}{\prod_{s=1}^{t} (1 + \widetilde{R}_s(\theta))}. \tag{24}$$

Minimiser $\mathbf{E}[\Lambda_\theta]$ revient à choisir l'allocation qui, en moyenne, « amortira » au mieux les flux de prestations futurs.

Notons que $L_0$ et $E_0$ étant fixés, minimiser $\mathbf{E}[\Lambda_\theta]$ revient également à maximiser la quantité $(E_0 + L_0 - \mathbf{E}[\Lambda_\theta])$ que l'on désignera dans la suite comme étant « les fonds propres économiques » de l'assureur en 0.

Il est donc possible d'écrire le programme d'optimisation suivant :

$$\begin{cases} \mathbf{Inf\ E}\left[\Lambda_\theta\right] \\ \theta \in [0;1] \end{cases} \qquad (25)$$

Par la suite, nous désignerons ce programme comme étant le critère de « maximisation des fonds propres économiques » (MFPE dans la suite). Les techniques de Monte Carlo permettent de résoudre ce problème :

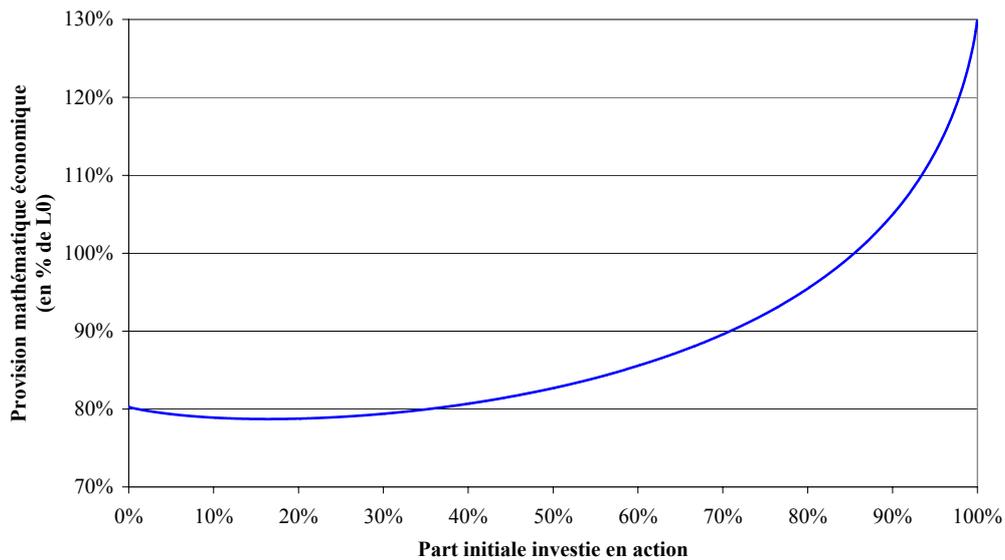

Fig. 5 :  Provision mathématique économique (en pourcentage de $L_0$) en fonction de la part d'actif risqué

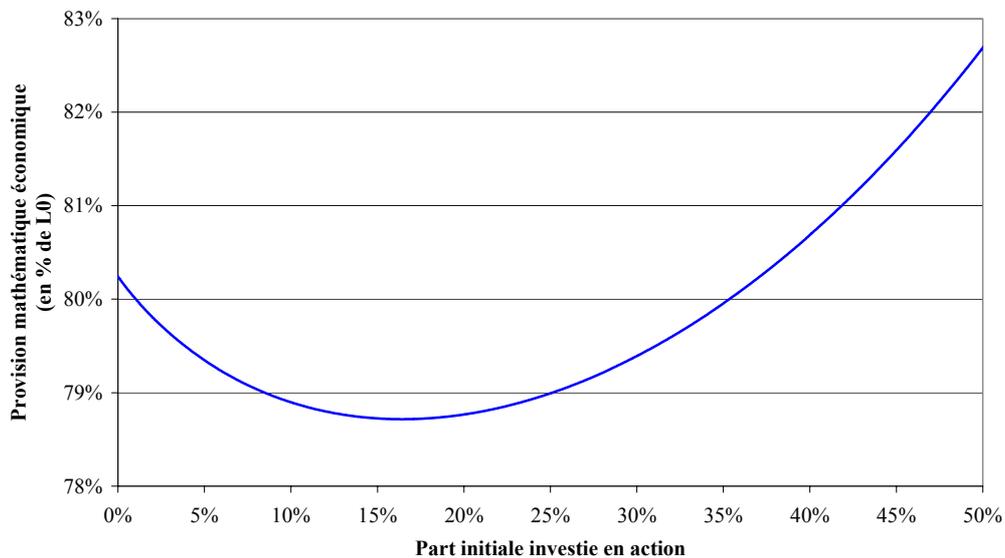

Fig. 6 :  Provision mathématique économique (en pourcentage de $L_0$) en fonction de la part d'actif risqué

Avec les mêmes paramètres que précédemment, le portefeuille optimal est composé pour 16,4 % d'actif risqué. Cette proportion initiale investie en actif risqué correspond à une probabilité de ruine comptable de 12,5 %. La ruine comptable intervient en moyenne pendant la deuxième année et son ampleur moyenne est de 372 k€ ce qui représente moins de 1 % de la provision mathématique initiale $L_0$. Néanmoins, pour cette allocation, aucune des simulations qui ont été effectuées n'a conduit à la « ruine économique » c'est à dire à l'épuisement total des ressources de l'assureur.

Par construction, ce critère d'allocation d'actifs induit une bonne adéquation des flux d'actif et de passif. Enfin, il présente l'avantage de ne pas avoir à fixer *a priori* un paramètre exogène tel que le niveau de la probabilité de ruine par exemple.

### 4.2.3. Etude de sensibilité

A l'image de la probabilité de ruine, l'allocation optimale selon le critère de maximisation des fonds propres économiques est relativement sensible aux paramètres des modèles d'actifs.

Les graphiques suivants présentent l'évolution de l'allocation optimale en fonction de la variation du paramètre étudiée et la probabilité de ruine associée à cette allocation optimale ; les autres paramètres restant constants.

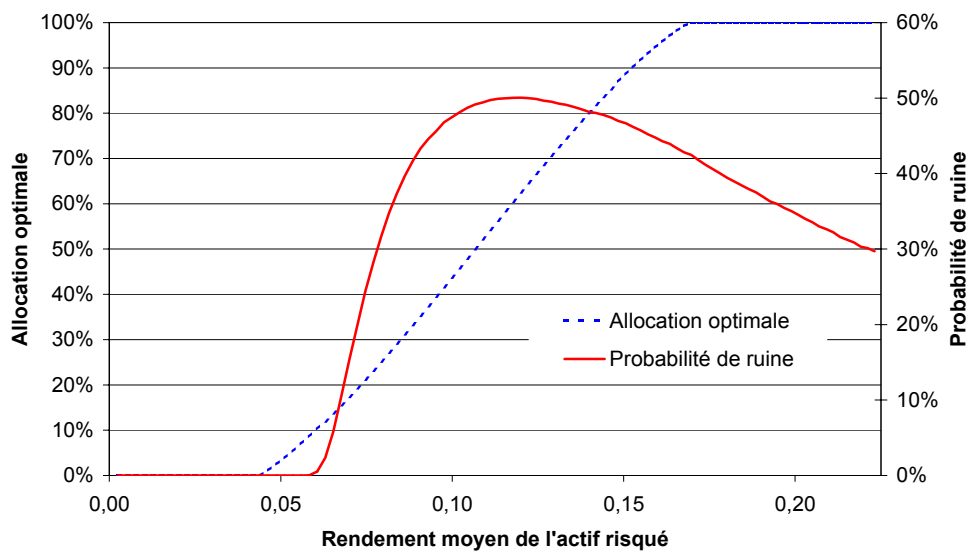

Fig. 7 :   Sensibilité au rendement de l'actif risqué

L'allocation optimale fournie par le critère de MFPE est évidemment croissante avec le rendement de l'actif risqué. En particulier, le portefeuille n'est diversifié que si $0,046 \leq \mu \leq 0,17$. La probabilité de ruine associée à l'allocation optimale est initialement croissante avec $\mu$ et donc avec l'allocation optimale puis décroissante pour $\mu \geq 0,12$.

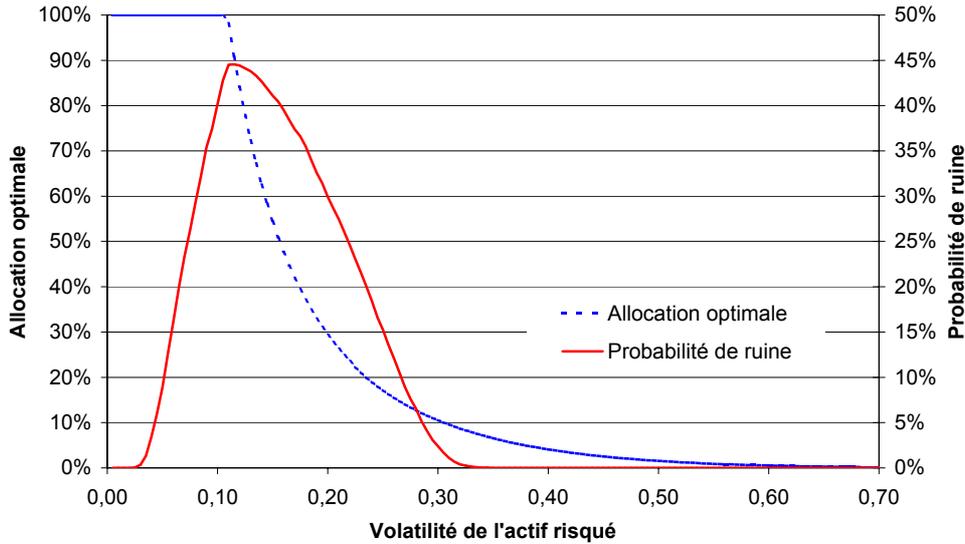

Fig. 8 : Sensibilité à la volatilité de l'actif risqué

L'allocation optimale fournie par le critère de MFPE est décroissante avec la volatilité de l'actif risqué. Le portefeuille est uniquement constitué d'actif risqué pour $\sigma_X \leq 0,1$. La probabilité de ruine est croissante dans un premier temps avec l'augmentation de la volatilité de l'actif risqué et le maintien de l'allocation optimale à des niveaux proches de 1 puis décroît avec l'allocation optimale lorsque la volatilité croît au delà de 0,11.

## 5. FLUX DE PRESTATIONS ALÉATOIRES

Dans ce paragraphe, les flux de prestations sont aléatoires : ils dépendent de l'évolution démographique du groupe de rentiers. Le graphique suivant présente l'intervalle de confiance à 95 % des prestations futures.

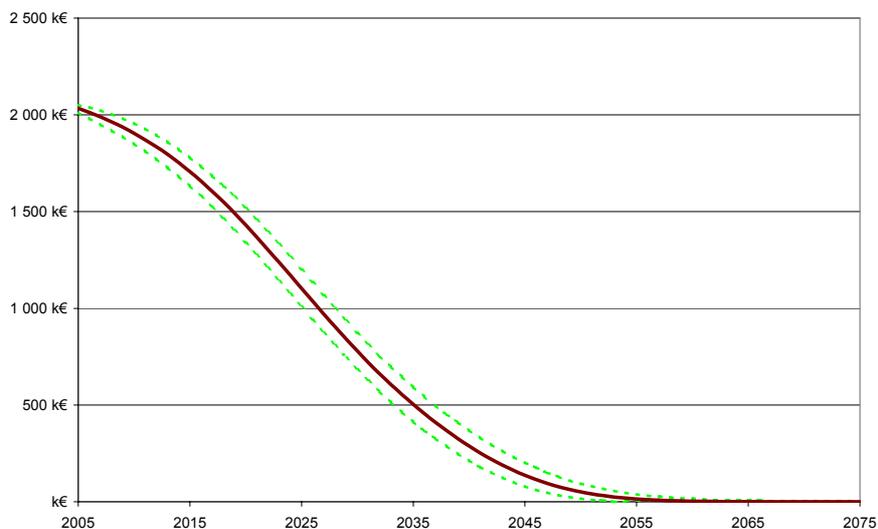

Fig. 9 : Intervalle de confiance à 95% des flux de prestations futurs

Cet intervalle est relativement étroit du fait de la bonne mutualisation des décès. Toutefois, comme le montrent CURRIE et al. [2004], le risque de mortalité comporte une autre composante, non mutualisable, associée à l'évolution aléatoire du taux instantané de décès autour de la tendance modélisée par les tables prospectives. De nombreux auteurs proposent des modélisations adaptées à la mesure de cette partie systématique du risque de mortalité : on pourra notamment se reporter à CAIRNS et al. [2004], DAHL [2004], LEE [2000] et SCHRAGER [2004].

L'intégration de ces approches au présent travail s'avère cependant délicate, car nous ne disposons pas d'un modèle fiable étalonné sur des données françaises qui pourrait être mis en œuvre ici. Nous nous restreindrons donc dans cette étude à la partie non systématique, et donc mutualisable, du risque de mortalité. En particulier, nous supposerons que la table de mortalité utilisée correspond parfaitement à la loi de probabilité de la mortalité des rentiers et nous étudierons l'impact des fluctuations d'échantillonnage sur l'allocation d'actifs.

Après avoir étudié la part du risque de mortalité dans le risque global, nous examinons l'impact de sa prise en compte sur l'allocation d'actifs.

## 5.1. ANALYSE DU RISQUE

Considérons la provision mathématique « économique » $\mathbf{E}[\Lambda_\theta]$. Lorsque les flux de prestations ne sont pas connus en 0, $\Lambda_\theta$ peut s'écrire :

$$\Lambda_\theta = \sum_{t=1}^{\infty} \frac{\widetilde{F}_t}{\theta X_t + (1-\theta) Y_t}. \tag{26}$$

La variance de $\Lambda_\theta$ peut être proposée comme indicateur du risque global du portefeuille. Sa décomposition nous permet d'apprécier les parts respectives des risques financiers et de mortalité. En effet, en conditionnant $\Lambda_\theta$ par l'évolution des placements, l'équation de décomposition de la variance conduit à l'égalité suivante :

$$\mathbf{V}[\Lambda_\theta] = \mathbf{E}[\mathbf{V}(\Lambda_\theta | X, Y)] + \mathbf{V}[\mathbf{E}(\Lambda_\theta | X, Y)]. \tag{27}$$

Le premier terme du membre de droite de l'expression (27) représente le risque financier associé au régime de rentes ; le second terme le risque technique de mortalité[10].

---

[10] Pour autant que la table de mortalité utilisée soit pertinente.

Via des techniques de simulations, en faisant l'hypothèse réaliste d'indépendance entre l'évolution des actifs et la mortalité, on génère $N$ trajectoires d'actifs risqués, $M$ trajectoires de passif et disposer ainsi de $N*M$ « états du monde » pour chaque θ.

Si $\lambda_{n,m}(\theta)$ est la réalisation de $\Lambda_\theta$ résultant de la $n$-ième trajectoire de l'actif risqué, de la $m$-ième trajectoire du passif, notons :

$$\overline{\lambda}_n(\theta) = \frac{1}{M}\sum_{m=1}^{M}\lambda_{n,m}(\theta), \tag{28}$$

et

$$\overline{\overline{\lambda}}(\theta) = \frac{1}{N}\sum_{n=1}^{N}\overline{\lambda}_n(\theta) = \frac{1}{N}\sum_{n=1}^{N}\frac{1}{M}\sum_{m=1}^{M}\lambda_{n,m}(\theta). \tag{29}$$

Les quantités (30) et (31) ci-après sont des estimateurs sans biais et convergents respectivement de $\mathbf{E}\left[\mathbf{V}(\Lambda_\theta|X,Y)\right]$ et de $\mathbf{V}\left[\mathbf{E}(\Lambda_\theta|X,Y)\right]$.

$$\frac{1}{N}\sum_{n=1}^{N}\frac{1}{M-1}\sum_{m=1}^{M}\left(\lambda_{n,m}(\theta) - \overline{\lambda}_n(\theta)\right)^2 \tag{30}$$

$$\frac{1}{N-1}\sum_{n=1}^{N}\left(\overline{\lambda}_n(\theta) - \overline{\overline{\lambda}}(\theta)\right)^2 \tag{31}$$

Ces résultats nous permettent d'obtenir le graphique 10 d'évolution du risque global associé au régime de rentes et d'évolution du risque financier dans le risque global en fonction du choix de portefeuille initial.

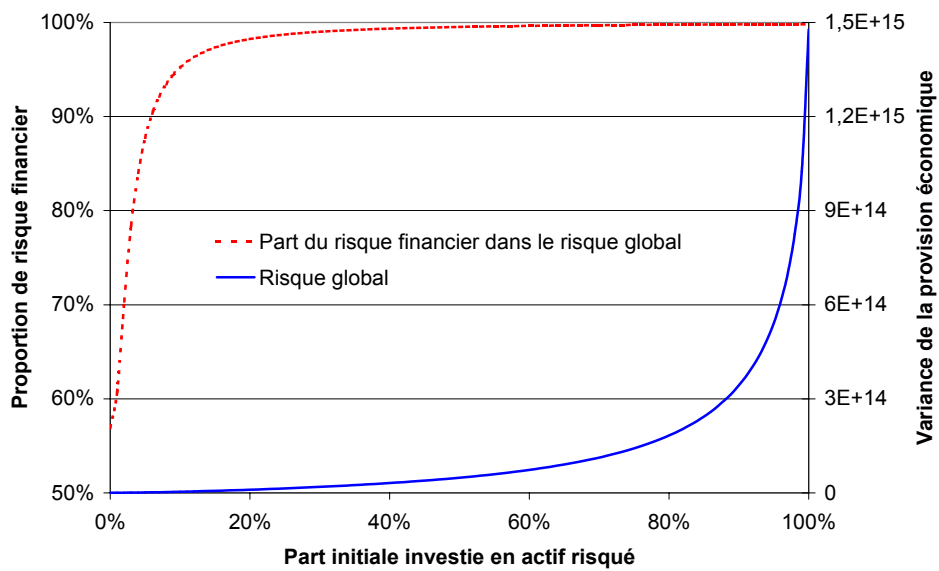

Fig. 10 : Evolution du risque global et de la part du risque financier dans le risque global en fonction de la part d'actif risqué

Le risque associé au régime est plus de 3 000 fois plus important lorsque l'actif est exclusivement composé d'actif risqué que lorsqu'il est uniquement composé de l'actif localement sans risque. Alors même que le nombre de rentiers est réduit et donc que la loi des grands nombres ne joue pas pleinement son rôle de mutualisation, la part du risque financier représente toujours plus 55 % du risque global. En particulier le risque de mortalité représente moins de 1 % du risque global dès lors que le portefeuille financier est initialement composé de plus de 30 % d'actif risqué.

Afin de mesurer l'impact de la mutualisation sur le risque supporté par le régime, nous avons répliqué à l'identique le portefeuille de rentiers de manière à ce qu'ils soient dix fois plus nombreux.

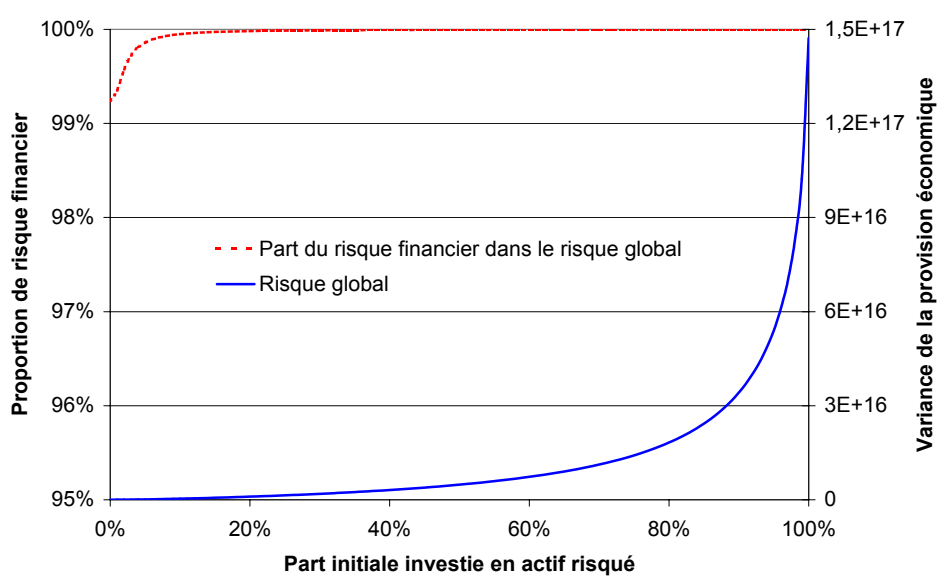

Fig. 11 : Evolution du risque global et de la part du risque financier dans le risque global en fonction de la part d'actif risqué pour un portefeuille de rentiers dix fois plus important

L'effet de la mutualisation s'observe par le fait que quelle que soit l'allocation d'actif effectuée, le risque global est toujours moins de 100 fois plus important lorsqu'il y a 3 740 individus que lorsqu'il y en a dix fois moins.

De ce fait, la part du risque financier dans le risque global est nettement plus importante que dans le cas précédent, ceci résulte du fait que le risque démographique est mutualisable et donc décroissant (en pourcentage) avec l'effectif alors que le risque financier est constant (en pourcentage) avec l'effectif. Le tableau suivant récapitule l'impact de la mutualisation sur l'allocation déterminée en 4.2.

|  | Part du risque financier dans le risque global | Part du risque financier dans le risque global (portefeuille x10) |
|---|---|---|
| $\theta = 16,4\ \%$ | 97,65 % | 99,98 % |

Tab. 1 : Effet de la mutualisation sur la part du risque financier

Ainsi lorsque le nombre de rentier est assez important et dans le cas d'une loi de mortalité déterministe, le risque purement « assurantiel » est négligeable par rapport au risque financier.

### 5.2. ALLOCATION D'ACTIFS

Dans cette partie, nous étudions l'impact de la prise en compte du risque de mortalité sur les critères d'allocation proposés dans le paragraphe 4.

#### 5.2.1. Probabilité de ruine

Si $e_k^{n,m}(\theta)$ est le niveau de fonds propres à la date $k$, dans l'état du monde $n$ x $m$ pour l'allocation $\theta$, l'expression *infra* est un estimateur convergent de la probabilité de ruine :

$$1 - \frac{1}{N}\sum_{n=1}^{N}\frac{1}{M}\sum_{m=1}^{M}\prod_{k\geq 1} I_{[0;\infty[}\left(e_k^{n,m}(\theta)\right) \tag{32}$$

Avec les mêmes paramètres que précédemment, nous obtenons la courbe suivante :

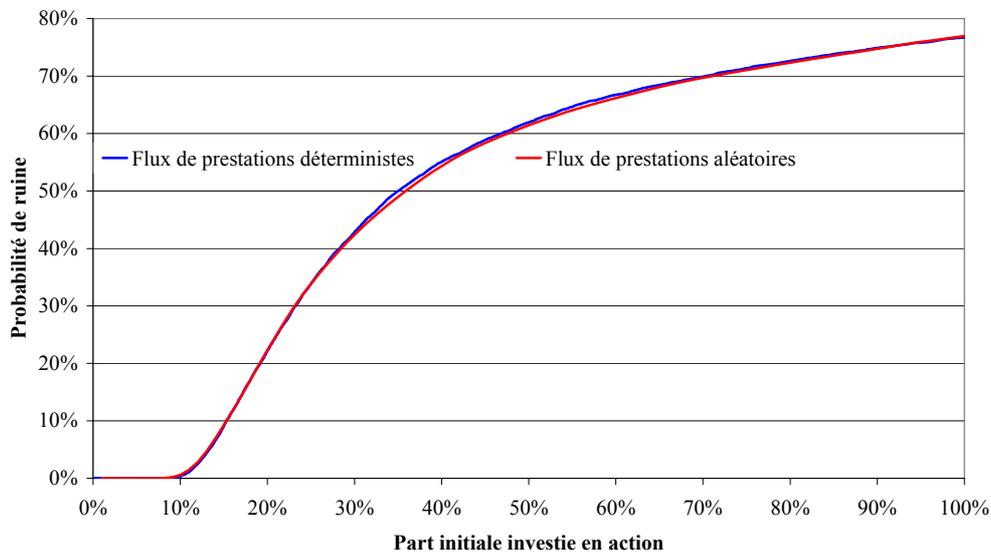

Fig. 12 : Probabilité de ruine en fonction de la part d'actif risqué

La courbe de la probabilité de ruine intégrant la nature aléatoire de la mortalité est pratiquement confondue avec celle obtenue en 4.1. Cela semble indiquer que la prise en compte de la nature aléatoire des flux de prestations affecte peu, dans le cas présent, l'allocation d'actifs déterminée à partir du critère de la probabilité de ruine.

Ce résultat est toutefois peu intuitif puisque la prise en compte de la mortalité augmente le risque global (au travers d'une augmentation de la variance), les effets dus aux variations de la mortalité se compensent donc en moyenne.

### 5.2.2. Maximisation des fonds propres économiques

Les flux de prestations et les rendements des actifs financiers étant indépendants, on a :

$$\mathbf{E}\left[\Lambda_\theta\right] = \mathbf{E}\left[\sum_{t=1}^{\infty} \frac{\widetilde{F}_t}{\theta X_t + (1-\theta)Y_t}\right] = \sum_{t=1}^{\infty} \mathbf{E}\left[\frac{1}{\theta X_t + (1-\theta)Y_t}\right] * \mathbf{E}\left[\widetilde{F}_t\right] \qquad (33)$$

Donc le programme d'optimisation qui consiste à maximiser le capital économique en 0 aura, lorsque les flux de prestations sont aléatoires, la même solution que lorsque ces flux sont connus en 0. Cette propriété est un avantage de ce critère puisque la prise en compte du risque de mortalité s'avère gourmande en temps de simulations.

### 5.2.3. Conclusion sur la prise en compte du risque de mortalité

Bien que non négligeable au vu du faible nombre de rentiers, le risque de mortalité n'affecte pas le choix de portefeuille au sens des deux critères exposés. Ce résultat est d'autant plus intéressant que la prise en compte du risque technique dans le critère de la probabilité de ruine demande des temps de calculs importants. En effet générer une simulation du passif nécessite de générer une trajectoire de mortalité pour chacun des 374 individus et de réévaluer le montant de la provision mathématique à chaque date en fonction des effectifs encore sous risque, le nombre de réalisations de variables aléatoires à simuler est donc très important.

## 6. REVALORISATION DES RENTES

Nous avons jusqu'à présent supposé que les rentes n'étaient pas revalorisées ; ceci n'est bien entendu pas réaliste, et il est nécessaire que le régime prévoie des règles de revalorisation des pensions. Les deux formes de revalorisations que l'on observe le plus fréquemment dans les régimes de pensions de retraite sont la revalorisation alignée sur l'évolution des salaires ou celle visant à maintenir le pouvoir d'achat des retraités. A long terme, le choix du dispositif n'est pas neutre, puisque pour autant que la croissance perdure, l'évolution des prix est généralement inférieure à la croissance des salaires. Dans ce paragraphe, nous supposerons que les rentes sont revalorisées de manière à tenir compte de l'inflation, pour maintenir le pouvoir d'achat des rentes[11].

---

[11] Ce qui se trouve être la situation des régimes de retraite obligatoires aujourd'hui.

Notons $I_t$ l'indice des prix à la consommation à la date $t$ et $L_t^I$ le montant de la provision mathématique à la date $s$. Cette quantité s'obtient[12] par :

$$L_s^I = \sum_{k=1}^{\infty} \frac{\mathbf{E}\left[\widetilde{F}_{s+k} * \frac{I_{s+k}}{I_s} \middle| \Phi_s\right]}{(1+i)^k}. \tag{34}$$

En faisant l'hypothèse réaliste d'indépendance entre l'évolution de la mortalité et celle des prix, il vient :

$$L_s^I = \sum_{k=1}^{\infty} \frac{\mathbf{E}\left[\widetilde{F}_{s+k} \middle| \Phi_s\right] * \mathbf{E}\left[\frac{I_{s+k}}{I_s} \middle| \Phi_s\right]}{(1+i)^k}. \tag{35}$$

La revalorisation des rentes représente un engagement supplémentaire pour l'assureur donc la provision mathématique de l'assureur sera plus importante que lorsque les rentes n'étaient pas revalorisées :

$$L_0^I \geq L_0. \tag{36}$$

La quantité $L_0^I - L_0$ peut alors s'interpréter comme la valeur en 0 de l'engagement de revaloriser les rentes de manière à maintenir le pouvoir d'achat des bénéficiaires.

### 6.1. MODÉLISATION DE L'INFLATION

De nombreux modèles ont été proposés dans la littérature pour rendre compte de l'inflation ; dans le cadre de modèles actuariels, on peut notamment citer les travaux de WILKIE [1995] et de KAUFMANN et al. [2001]. Nous retenons ici la modélisation proposée par AHLGRIM et al. [2003] qui suppose que l'indice des prix à la consommation évolue de la manière suivante :

$$I_{t+\delta} = I_t * \exp \int_t^{t+\delta} (j + x_s) ds, \tag{37}$$

où $j$ est le taux instantané moyen d'évolution des prix et où le processus $x$ modélise les fluctuations aléatoires autour de la tendance :

$$dx_s = -a\, x_s\, ds + \sigma_I\, dB_s^I, \tag{38}$$

---
[12] Le fait d'introduire un processus de revalorisation des rentes lié à l'évolution de l'inflation ne modifie pas l'approche de valorisation des flux futurs probables sous la probabilité historique retenue dans cet article.

où $B^I$ un **P**-mouvement brownien. Nous ferons l'hypothèse d'indépendance entre le processus $x$ et l'évolution des taux d'intérêt ; en effet, la modélisation que nous retenons de l'inflation incorpore une tendance déterministe, avec des fluctuations aléatoires autour ; une éventuelle dépendance entre $x$ et le niveau des taux d'intérêt est donc ici un effet du second ordre, que nous négligerons.

On cherche donc à calculer la quantité :

$$\mathbf{E}\left[\frac{I_{t+\delta}}{I_t}\bigg|\Phi_t\right] = \mathbf{E}\left[\exp\int_t^{t+\delta}(j+x_s)ds\bigg|\Phi_t\right] = e^{j\delta}\mathbf{E}\left[\exp\int_t^{t+\delta}x_s\,ds\bigg|\Phi_t\right]. \tag{39}$$

Puisque $x$ est un processus d'Ornstein-Ulhenbeck, la variable aléatoire $x_s$ est une gaussienne dont on peut calculer moyenne et variance :

$$\begin{aligned}\mathbf{E}[x_s] &= x_0 e^{-as} \\ \mathbf{V}[x_s] &= \sigma_I^2\frac{1-e^{-2as}}{2a}\end{aligned} \tag{40}$$

De plus $\int_t^{t+\delta}x_s\,ds$ est une gaussienne puisqu'il s'agit de la limite d'une somme de variables aléatoires gaussiennes. L'expression de la transformée de Laplace d'une gaussienne[13] nous permet donc d'écrire :

$$\mathbf{E}\left[\exp\int_t^{t+\delta}x_s\,ds\bigg|\Phi_t\right] = \exp\left\{\mathbf{E}\left[\int_t^{t+\delta}x_s\,ds\bigg|\Phi_t\right] + \frac{1}{2}\mathbf{V}\left[\int_t^{t+\delta}x_s\,ds\bigg|\Phi_t\right]\right\}. \tag{41}$$

Comme[14]

$$\mathbf{E}\left[\int_t^{t+\delta}x_s\,ds\bigg|\Phi_t\right] = x_t\frac{1-e^{-a\delta}}{a}, \tag{42}$$

et :

$$\mathbf{V}\left[\int_t^{t+\delta}x_s\,ds\bigg|\Phi_t\right] = \frac{\sigma_I^2}{a^2}\left(\delta - \frac{1-e^{-a\delta}}{a} - \frac{(1-e^{-a\delta})^2}{2a}\right), \tag{43}$$

il vient :

$$\mathbf{E}\left[\exp\int_t^{t+\delta}x_s\,ds\bigg|\Phi_t\right] = \exp\left\{x_t\frac{1-e^{-a\delta}}{a} + \frac{\sigma_I^2}{2a^2}\left(\delta - \frac{1-e^{-a\delta}}{a} - \frac{(1-e^{-a\delta})^2}{2a}\right)\right\}, \tag{44}$$

---

[13] Rappelons que si $\varepsilon$ est une variable aléatoire de loi $N(m,\sigma^2)$, $\mathbf{E}[\exp(z*\varepsilon)] = \exp\left\{zm + z^2\frac{\sigma^2}{2}\right\}$.

[14] *Cf.* LAMBERTON et al. [1997] pour le détail des calculs.

et enfin :

$$\mathbf{E}\left[\frac{I_{t+\delta}}{I_t}\bigg|\Phi_t\right]=\exp\left\{j\delta+x_t\frac{1-e^{-a\delta}}{a}+\frac{\sigma_I^2}{2a^2}\left(\delta-\frac{1-e^{-a\delta}}{a}-\frac{(1-e^{-a\delta})^2}{2a}\right)\right\}. \qquad (45)$$

Le nouveau montant de la provision mathématique en 0 est donc donné par :

$$L_0^I = \sum_{k=1}^{\infty}\frac{\mathbf{E}[\widetilde{F}_k|\Phi_0]*\exp\left\{jk+x_0\frac{1-e^{-ak}}{a}+\frac{\sigma_I^2}{2a^2}\left(k-\frac{1-e^{-ak}}{a}-\frac{(1-e^{-ak})^2}{2a}\right)\right\}}{(1+i)^k}. \qquad (46)$$

Nous retiendrons pour les applications numériques, les paramètres suivants estimés par FARGEON et al. [2003] sur les données de l'inflation française depuis janvier 1990, soit :

$$j = 0{,}0279 \qquad a = 0{,}7369 \qquad \sigma_I = 0{,}0056$$

On obtient alors un montant de provision mathématique initiale de 47,9 M€. Le processus de revalorisation des rentes conduit donc à augmenter l'engagement de l'assureur de 46 %. Le poids relatif de la revalorisation est donc très important. L'assureur devra augmenter d'autant la part de ses fonds propres alloués au régime. Par ailleurs les flux espérés en 0 ont l'allure suivante :

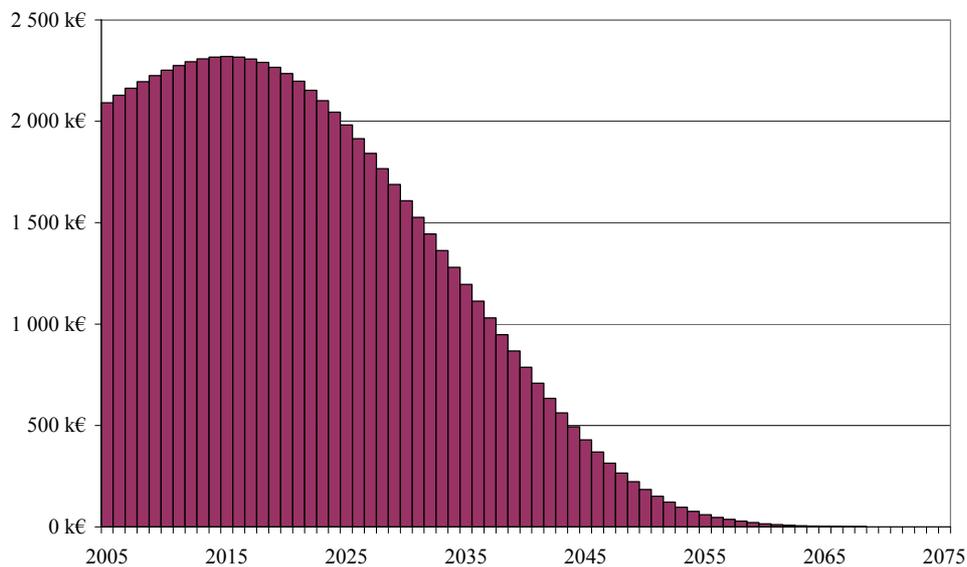

Fig. 13 : <u>Flux de prestations revalorisés espérés</u>

La courbe des flux futurs n'est plus strictement décroissante comme dans la cas où il n'y avait pas de revalorisation. En effet c'est l'année 2015 qui, en moyenne, verra le débours le plus important. La duration du passif augmente sensiblement pour s'établir à 14,9 ans.

Même avec une volatilité de l'inflation très petite $(\sigma_I = 0{,}0056)$, cette suite de flux est volatile puisque l'intervalle de confiance autour de cette courbe est très large en comparaison avec celui obtenu lorsque les rentes ne sont pas revalorisées.

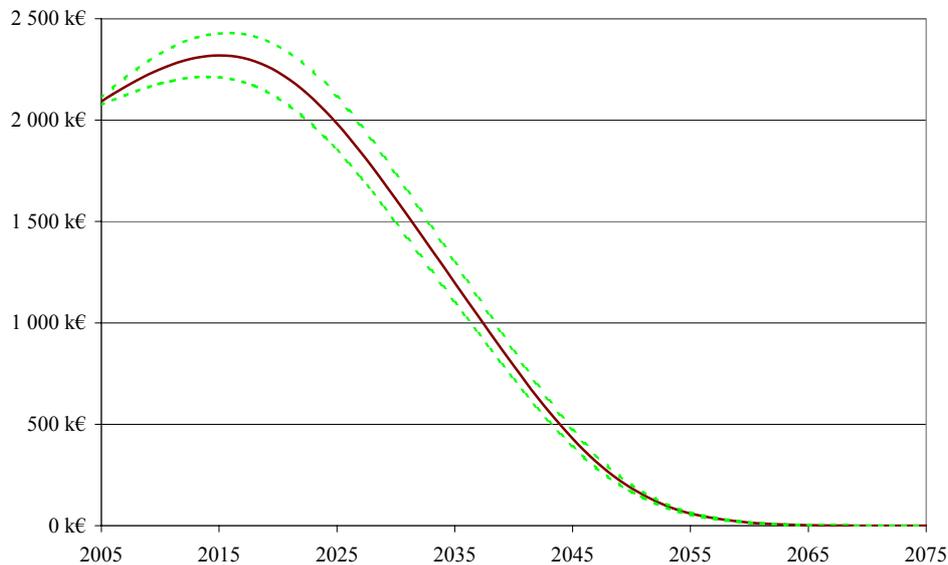

Fig. 14 :   Flux de prestations revalorisées – intervalle de confiance à 95%

La largeur de cet intervalle provient du fait que le risque d'inflation n'est pas mutualisable.

### 6.2. ALLOCATION D'ACTIFS

Dans cette partie, nous ferons l'hypothèse de parfaite mutualisation des décès et donc que les flux de passif sont connus en 0. En effet les illustrations du paragraphe 5 nous ont permis d'observer que la mortalité n'a pas d'impact sur l'allocation déterminée par le critère des fonds propres économiques et un impact négligeable sur l'allocation déterminée à partir de la probabilité de ruine.

#### 6.2.1. Probabilité de ruine

Le graphique suivant compare les probabilités de ruine selon que les rentes sont ou non revalorisées.

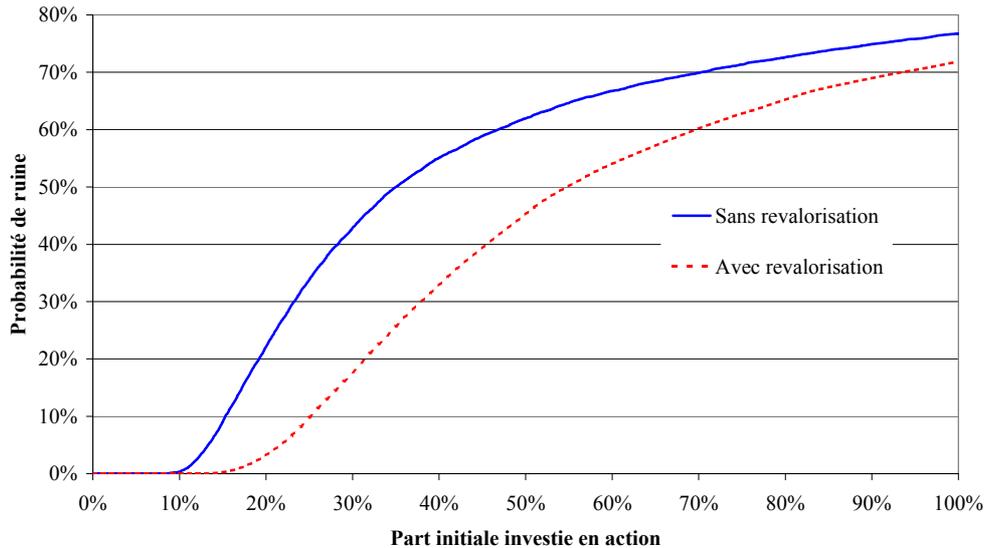

Fig. 15 : Probabilité de ruine en fonction de la part d'actif risqué

Quelle que soit l'allocation en actif risqué, la probabilité de ruine est inférieure à celle obtenue lorsque les rentes n'étaient pas revalorisées. En effet, bien que l'assureur supporte le risque supplémentaire de voir l'inflation augmenter de manière importante et de ne pas disposer d'assez de réserves pour faire face à ses engagements, l'augmentation de la duration du passif et par là de la durée de vie moyenne de l'actif joue en faveur de l'investissement en actif risqué. Ainsi, si l'on se fonde sur la probabilité de ruine pour choisir son allocation d'actifs, l'augmentation de la durée du passif est un facteur favorisant l'investissement en actif risqué.

6.2.2. Maximisation des fonds propres économiques

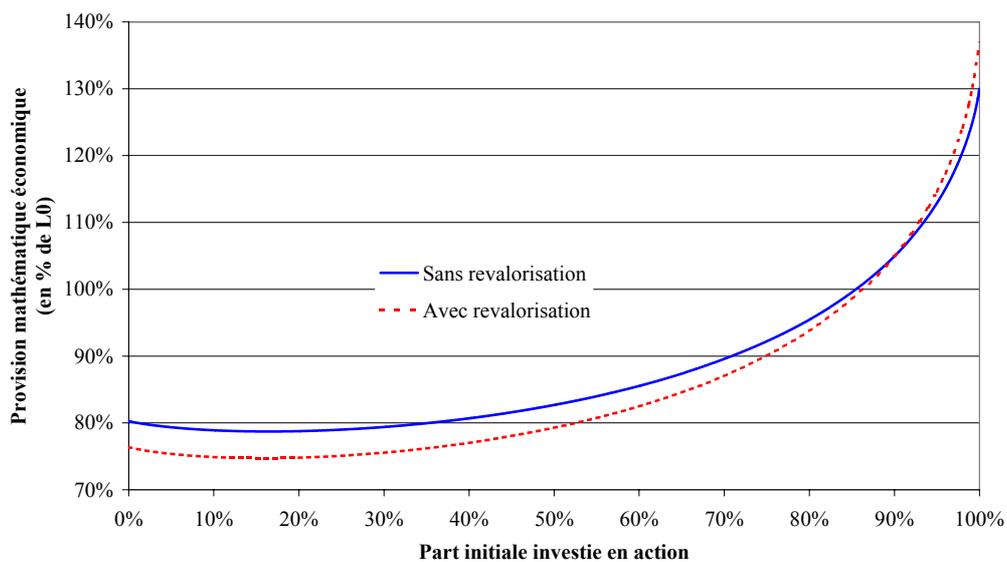

Fig. 16 : Provision mathématique économique (en pourcentage de $L_0$) en fonction de la part d'actif risqué

Comme l'indique le graphique 16 la provision mathématique économique exprimée en fonction de la provision initiale admet un minimum non trivial pour θ = 0,158.

## 7. CONCLUSION

L'application des techniques de simulation à des portefeuilles réels permet de proposer des allocations stratégiques d'actifs intégrant des critères plus complexes, et sans doute mieux adaptés à l'activité de l'assurance vie, que le traditionnel modèle de Markowitz et ses dérivés.

Le présent article propose, pour un régime de rentiers, un critère d'allocation s'appuyant sur la minimisation de l'espérance des flux de prestations futurs actualisés au taux de rendement de l'actif de la société. Ce critère (MFPE) est illustré sur un portefeuille réel et fournit une allocation relativement prudente par rapport aux pratiques du marché. A l'image de critères tels que la probabilité de ruine, il s'avère relativement sensible aux paramètres des modèles d'actifs utilisés.

Le critère de MFPE présente toutefois deux avantages : tout d'abord, il intègre la structure du passif par le biais du profil de flux futurs et ne nécessite pas la simulation de la mortalité lorsque l'on fait l'hypothèse d'indépendance entre celle ci et l'évolution des marchés financiers et de l'inflation. En termes opérationnels, ce dernier point est important, puisque la mortalité est lourde à simuler même sur un groupe très réduit. Par ailleurs la prise en compte de l'obligation réglementaire de revaloriser les rentes ne pose pas de problème pratique.

De plus, l'utilisation de ce critère ne repose pas sur un paramètre exogène, tel que le niveau de la probabilité de ruine, dont la détermination peut prêter à discussion (*cf.* les travaux en cours concernant Solvabilité 2) ou encore sur le choix délicat d'une fonction d'utilité de l'assureur.

L'objectif du présent travail était de valider la pertinence du critère MFPE dans le cas d'un régime de rentiers dans un cadre simplifié ; nous poursuivons nos travaux pour intégrer au modèle la prise en compte d'obligations indexées sur l'inflation (OATi), de la mortalité stochastique (partie non mutualisable du risque de mortalité) et la politique de rémunération des actionnaires.

# BIBLIOGRAPHIE